\documentclass[prb,twocolumn,showpacs,groupedaddress]{revtex4}

\usepackage{graphicx}

\font\grg=eurm10
\def\umu{{\hbox{\grg\char22}}}
\font\grs=eurm10 at 9pt
\def\smu{{\hbox{\grs\char22}}}

\begin{document}

\bibliographystyle{apsrev}

\title{Magnetic Field Dependent Coherent Polarization Echoes in Glasses}

\author{S. Ludwig$^*$, P. Nagel, S. Hunklinger, and
C. Enss} \affiliation{Kirchhoff-Institut f\"ur Physik, Universit\"at
Heidelberg, INF 227, 69120 Heidelberg, Germany}
\date{\today}

\begin{abstract}
The unexpected finding of a strong magnetic field dependence of the
dielectric properties of insulating glasses at very low temperatures has
been a puzzling problem since its discovery. Several attempts have been
made to explain this striking phenomenon. In order to obtain information
on the origin of the magnetic field effects we have studied coherent
properties of atomic tunneling systems in glasses by polarization echo
experiments in magnetic fields. Our results clearly favor a model based on
the assumption that nuclear quadrupoles play a crucial role for the
observed magnetic field dependence.\\
$^*$present address: Department of Physics, Stanford University, Stanford,
CA 94305-4060, USA

\end{abstract}


\maketitle

\section{Introduction}

The absence of first order magnetic field effects in insulating materials
free of magnetic moments can be concluded from basic principles. Since the
magnetic field is described by an axial vector and the electrical
polarization by a polar vector, a linear magnetic field dependence of the
dielectric properties of pure non-magnetic insulators can be ruled out.
Therefore, it is very surprising that strong magnetic field effects were
discovered recently in low-frequency dielectric susceptibility
measurements on certain multi-component glasses at very low
temperatures.\cite{Str98,Str00,Woh01,Hau01,Coc02} In particular, a
non-monotonic variation of the dielectric constant and of the dielectric
loss was observed with increasing magnetic field.

Since the low-temperature dielectric properties of insulating glasses are
governed by atomic tunneling systems it was speculated on a direct
coupling of magnetic fields to tunneling systems. Two models have been
proposed that relate the magnetic field dependence to the Ahanorov-Bohm
phase of a charged particle moving along a closed loop. \cite{Ket99,Wue02}

In polarization echo experiments at radio frequencies it has already been
demonstrated that tunneling systems in glasses couple directly to magnetic
fields.\cite{Lud02} Surprisingly, the amplitude of two-pulse echoes in
a-BaO-Al$_2$O$_3$-SiO$_2$ (BAS) was found to depend strongly on the
applied magnetic field showing a non-monotonic field variation. Since in
polarization echo experiments only the properties of tunneling states are
probed these findings prove that indeed the tunneling systems themselves
are involved in the magnetic field effect. Furthermore, the character of
certain features observed in these echo experiments indicate that nuclear
magnetic moments might play an important role.

To explain the magnetic field effects in echo experiments a coupling to
nuclear spins has recently been discussed theoretically.\cite{Wue02b} The
model worked out in this publication is based on the assumption that the
levels of tunneling particles with non-zero nuclear quadrupole moment
exhibit a quadrupole splitting, which is different in the ground state and
in the excited state. Magnetic fields cause an additional Zeeman splitting
of these levels giving rise to interference effects. In turn, these
effects cause the non-monotonic magnetic field variation of the echo
amplitude.

In this communication we present the results of polarization echo
experiments on four different glasses in magnetic fields investigated
under different experimental conditions. Furthermore, we discuss briefly
the models developed to explain the magnetic field effects and compare
their predictions with our experimental findings.

\section{Theoretical Background}

In the following sections we briefly introduce the tunneling model,
which is the generally accepted basis for the description of the
low-temperature properties of amorphous solids. In addition, we discuss
the physics of polarization echoes in glasses, and the theoretical
approaches to explain the magnetic field dependence of the dielectric
properties of non-magnetic insulating glasses.
\bigskip\bigskip

\centerline{\bf A. Tunneling Model}
\bigskip

Atomic tunneling states give rise to an important contribution to the
internal energy of glasses at low temperatures. Most properties of
amorphous solids are strongly influenced by these additional degrees of
freedom. Two prominent examples are the linear specific heat and the
logarithmic temperature dependence of the sound
velocity.\cite{Esq98,Hun00} A phenomenological description of the
low-temperature properties of glasses is provided by the so-called \lq
tunneling model', that has been proposed independently by Phillips
\cite{Phi72} and Anderson et al.\cite{And72} in 1972. A central assumption
of this model is that atoms or small groups of atoms are not located in
well-defined potential minima, but move between two energetically almost
equivalent adjacent positions separated by a potential barrier. Such a
configuration can be described by a \lq particle' moving in a double-well
potential. In this approximation the single wells are considered as
harmonic and identical, but may differ in their depth. This difference is
usually referred to as the asymmetry energy~${\it\Delta}$. These particles
have vibrational states in each single well, separated by the
energy~$E_0$.

At low temperatures, i.e., for $k_{\rm B}T \ll E_0$, higher vibrational
levels are not excited and can be omitted in our further discussion. The
ground state is split into two levels by the tunneling motion. Therefore,
tunneling systems in glasses are often referred to as {\em two-level
systems}. In addition to the classical potential difference~${\it\Delta}$,
the quantum mechanical tunnel splitting~${\it\Delta}_0$ contributes to the
ground state splitting~$E$ given by
\begin{equation}
E=\sqrt{{\it\Delta}^2+{\it\Delta}_0^2}\; . \label{eq1}
\end{equation}
\noindent Applying the WKB approximation the tunnel splitting can be
calculated approximately. Neglecting prefactors of the order of unity one
finds
\begin{equation}
{\it\Delta}_0 \approx E_0\, {\rm e}^{-\lambda}\;. \label{eq2}
\end{equation}

Roughly speaking the tunnel splitting~${\it\Delta}_0$ is given by the
vibrational energy~$E_0$ of the particle multiplied by the probability
$\exp\, (-\lambda)$ for tunneling. The so-called tunneling parameter
$\lambda = d\sqrt{2mV}/2\hbar$ reflects the overlap of the wave functions
of the particle at the two sides of the potential barrier.

As a consequence of the irregular structure of glasses, the characteristic
parameters of the tunneling states are widely distributed. In the
tunneling model it is assumed that the asymmetry energy~${\it\Delta}$ and
the tunnel parameter~$\lambda$ are independent of each other and uniformly
distributed as
\begin{equation}
P(\lambda,{\it\Delta})\,{\rm d}\lambda\,{\rm d}{\it\Delta} =
\overline{P}\,{\rm d}\lambda\,{\rm d}{\it\Delta} \; , \label{eq3}
\end{equation}
\noindent where $\overline{P}$ is a constant. Using this particular
distribution function many low-temperature properties of glasses can be
explained even quantitatively. An equivalent description of the
low-temperature properties of glasses is given by the so-called \lq soft
potential model' \cite{Kar83} that uses a more general form of the atomic
potentials and different distribution functions. It allows reliable
predictions even well above 1\,K. In comparison with the tunneling model
the soft potential model leads to a slightly different distribution of the
tunneling parameter, but for the properties below 1\,K it yields the same
predictions.

Tunneling systems couple to their environment by interaction with phonons
and photons. External elastic or electric fields produce changes of the
asymmetry energy~${\it \Delta}$ and lead to relaxation processes. The
coupling to external fields also causes resonant processes like resonant
absorption and stimulated emission. Both, resonant and relaxational
contributions determine the dynamic response of the two-level systems. For
a tunneling system with electric dipole moment~$\bf p$ in an electric
field~${\bf F}(t)$ the Hamiltonian in the orthogonal basis of the
eigenfuntions~$\psi_-$ and~$\psi_+$ of ground and excited state is given
by
\begin{eqnarray}
\!\!H={1\over 2}\pmatrix{ E & 0\cr 0 & -E\cr}\;+\;{1\over E}
\pmatrix{{\it\Delta}   & -{\it\Delta}_0     \cr
         -{\it\Delta}_0 & -{\it\Delta}       \cr}
        \;{\bf p}\cdot{\bf F}(t)
\end{eqnarray}
Using this Hamiltonian the coherent dynamics of tunneling systems can be
formulated.
\bigskip\bigskip

\centerline{\bf B. Polarization Echoes in Glasses}
\bigskip

At very low temperatures the relaxation of the tunneling systems becomes
so slow that coherent phenomena like polarization echoes become observable
in insulating glasses. Depending on the sequence and phase of the exciting
pulses, different physical properties of the tunneling systems can be
studied.\cite{Gol81,Ens96} Here we focus on one particular kind of
experiment, the so-called spontaneous echo (or two-pulse echo) which is
generated by two short microwave pulses separated by the delay time
$t_{12}$. The macroscopic polarization produced by the first microwave
pulse vanishes rapidly due to the distribution of parameters of the
tunneling systems in glasses. This phenomenon is similar to the well-known
free induction decay observed in nuclear magnetic resonance experiments.

Neglecting relaxation processes and spectral diffusion\cite{Bla77}
resulting from the interaction of neighbored tunneling systems, the phase
of the individual tunneling systems develops freely between the two
exciting pulses. The second pulse causes an effective time reversal of the
development of the phase of the tunneling systems. The initial macroscopic
polarization of the glass is recovered at the time~$t_{12}$ after the
second pulse. In analogy to the two-pulse echo in magnetic resonance
experiments this phenomenon is referred to as the spontaneous echo. Due to
the broad distribution of the parameters of the two-level systems in
glasses, the description of polarization echoes in glassy materials is
somewhat more involved than in the simple case of diluted paramagnetic
spin systems. A detailed and general theoretical discussion of spontaneous
echoes in glasses has been given by Gurevich et al. \cite{Gur90}. A few
results of this theory relevant to the experiments presented here, will be
summarized in the following.

The amplitude of the echo depends on the pulse area, i.e., the length of
the exciting pulses times the field strength of the microwave. A maximum
of the echo amplitude is observed for pulses with pulse areas
${\it\Omega}t_{\rm p1}\approx \pi/2$ and ${\it\Omega}t_{\rm p2}\approx
\pi$, where $t_{\rm p1}$ and $t_{\rm p2}$ denote the duration of the first
and second pulse, respectively, and ${\it\Omega}$ the effective Rabi
frequency given by
\begin{equation}
{\it\Omega}=\sqrt{{\it\Omega}_{\rm R}^2+\omega_{\rm d}^2}\; .
\label{effective_Rabi_frequency}
\end{equation}
\noindent Here $\omega_{\rm d}=\omega_{\rm r}-\omega$ signifies the
difference between the frequency~$\omega$ of the microwave and the
resonance frequency~$\omega_{\rm r} = E/\hbar$ of the tunneling systems.
The quantity ${\it\Omega}_{\rm R}$ represents the Rabi frequency of
systems at $\omega_{\rm d}=0$ and is given by
\begin{equation}
{\it\Omega}_{\rm R}={1\over\hbar}{{\it\Delta}_0\over E}\,{\mathbf
p}\cdot{\mathbf F}_0\; , \label{Rabi_frequency}
\end{equation}
\noindent where $\vert{\mathbf F}_0\vert$ is the field strength of
the exciting microwave pulse. Neglecting again any kind of phase
disturbing processes, the amplitude of spontaneous echoes in
glasses is expected to vary as \cite{Gur90}
\begin{eqnarray}
A(t)&\!\!=\!\!&A_0\tanh\left({E\over 2k_{\rm B}T}\right)\,{\rm
Im}\Biggl\{{\rm e}^{{\rm i}\omega
t}\!\!\int\limits_0^1\!\cos^4\!\!{\it\Theta}\;{\rm
d}\!{\it\Theta}\,\int\limits_0^1\!{\eta\,{\rm d}\eta\over
\sqrt{1-\eta}}\nonumber\\
&&\int\limits_{-\infty}^{+\infty}{1\over
{\it\Omega}^3}\left[\sin\left({\it\Omega}t_{\rm p1}\right)-2{\rm
i} {\omega_{\rm d}\over
{\it\Omega}}\sin^2\!\!\left({{\it\Omega}t_{\rm p1}\over
2}\right)\right]\nonumber\\&&\sin\!\left({{\it\Omega}t_{\rm
p2}\over 2}\right)\,{\rm e}^{{\rm i}\omega_{\rm
d}(t-2t_{12})}\,{\rm d}\omega_{\rm d}\Biggr\} \; ,
\label{echo_amplitude}
\end{eqnarray}
\noindent with $\eta={\Delta_0/E}$, and ${\it\Theta}$ representing the
angle between ${\mathbf p}$ and ${\mathbf F}_0$. The amplitude $A_0$ is
determined by the number of tunneling states, their dipole moment and the
field strength of the microwave.

At finite temperatures, thermal relaxation processes and spectral
diffusion weaken the coherence of the ensemble of resonant tunneling
systems. Thus the echo amplitude is increasingly reduced with growing
delay time~$t_{12}$. Since thermal relaxation processes and spectral
diffusion are strongly temperature dependent, coherent polarization echoes
in glasses can be observed only at very low temperatures, typically below
100\,mK. We shall not discuss these processes here since they are not of
importance for the phenomena reported here. However, we like to point out
that the echo amplitude is proportional to the number of tunneling systems
which are in resonance with the exciting microwave pulse and do not loose
their phase coherence during the time $2t_{12}$.
\bigskip\bigskip

\centerline{\bf C. Coupling to Magnetic Fields}
\bigskip

Recently, several models have been developed with the aim to explain the
coupling of magnetic fields to atomic tunneling systems in disordered
solids. Here we discuss first two models that focus on the description of
the dielectric constant of glasses in magnetic fields. Then we consider
the model that aims to explain the magnetic field effects observed in
polarization echo experiments.
\bigskip

\centerline{\it 1. Mexican-hat Model}
\bigskip

To describe the non-monotonic magnetic field dependence of the dielectric
susceptibility of multi-component glasses, Kettemann et al.\cite{Ket99}
investigated the properties of tunneling systems exhibiting the
peculiarity that the tunneling particle can move along different paths to
go from one potential minimum to the other thus forming a closed tunneling
loop. As a simple example they considered a Mexican-hat type potential
with two equivalent minima in the brim. At zero magnetic field the
tunneling probability is equal for either direction of tunneling. However,
in a magnetic field perpendicular to the plane of the tunneling loop the
time inversion symmetry is broken and the probability for tunneling is in
general not equal anymore for the two directions of revolution. As a
result the tunnel splitting~${\it\Delta}_{0}$ becomes a periodic function
of the magnetic flux $\phi$ threading through the loop, i.e., \hbox{${\it
\Delta}_0(\phi)={\it\Delta}_{0}(\phi=0) \cos(\pi\phi/\phi_0)$}, where
$\phi_0=h/e$ represents the elementary flux quantum. Consequently, the
energy spectrum of the tunneling systems varies periodically with the
applied magnetic field and so does the dielectric constant of the glass.

In this model, maxima in the dielectric constant are expected at
$\phi/\phi_0=(2n+1)/2$, with $n= 0, 1, 2,\;\dots$. The magnetic field
required to reach the first maximum of the dielectric constant turns out
to be of the order of 10$^5$\,T if the tunneling systems carry an
elementary charge and have atomic dimensions. However, experimentally the
first maximum was found at about 100~mT. This clear contradiction was
interpreted by Kettemann et al. as an indication that not individual
tunneling systems are responsible for the magnetic field effect but a
large number of strongly coupled systems with a much larger effective
charge and a larger effective loop radius. Although evidence for
interactions between tunneling systems in glasses have been found in
several experiments at very low
temperatures\cite{Sal94,Rog96,Osh96,Bur98,Ens97,Ens02} it remains an open
question whether clusters of 10$^5$ coherently moving tunneling systems
exist in glasses.

No specific predictions were made for echo experiments in this work.
However, the predicted periodical variation of the tunnel splitting
should be observable since ${\it\Delta}_0$ enters in several
quantities relevant in echo experiments. For example, one such
quantity is the Rabi frequency.
\bigskip

\centerline{\it 2. Pair Model}
\bigskip

A somewhat different point of view had been taken by W\"urger\cite{Wue02},
who considered the properties of pairs of interacting tunneling systems in
magnetic fields. Suppose that there are two neighbored interacting
tunneling systems of similar energy with tunneling paths along different
directions. In this case the interaction between the tunneling particles
would lead to a bending of the tunneling paths and to a correlation of the
tunnel motion of the two systems. It is conceivable that a closed loop is
formed along which the effective charge of the tunneling systems is
transported. The important difference to the Mexican-hat model is that the
level scheme of a coupled pair differs from that of a two-level system.
Two weakly coupled two-level systems of similar energy splitting have four
levels, the two in the middle being almost degenerate. The separation of
the two levels in the middle will periodically vary as a function of
magnetic flux through the effective loop.

Within this approach, the splitting of the two almost degenerate levels is
given by the relation $E_{\rm p}=\sqrt{\delta^2+\zeta^2}$, with $\delta$
representing the difference in the energy splitting of the two coupled
tunneling systems $\delta=E_1-E_2$, and $\zeta$ denoting their magnetic
field dependent variation
\hbox{$\zeta=2{\it\Delta}_0\sin(\pi\phi/\phi_0)$}. The important feature
of this model is that the relative change of the small splitting $E_{\rm
p}$ caused by magnetic fields is much larger than the relative variation
of the energy splittings~$E_1$ and~$E_2$ of isolated tunneling systems.
Within this model the magnetic field variation~$\Delta\varepsilon$ of the
dielectric constant~$\varepsilon$ is approximately given by
\begin{equation}
\Delta\varepsilon\;\propto\;{1\over
B}\left[1-\cos\left({2\pi{\it\Gamma}\,\phi^2\over
\phi_0^2}\right)\right]\; , \label{DK_Pairs}
\end{equation}

\noindent with ${\it\Gamma}= 2\pi{\it\Delta}_0^2/(\hbar\omega p F)$. Here,
$B$ denotes the magnetic field, $\omega$ the frequency of measurement and
$F$ the amplitude of the applied ac field. Inserting typical values of
low-frequency measurements one can estimate the parameter~${\it\Gamma}$ to
be ${\it\Gamma}\approx 10^{10}$. Therefore, the oscillation period is
determined by an effective flux quantum $\phi_0/\sqrt{\it\Gamma}\approx
10^{-5}\phi_0$, which appears to be of the right order of magnitude for
describing the observed non-monotonic magnetic field variation of
$\varepsilon$. However, it should be pointed out, that this explanation
relies unambiguously on the presence of coupled pairs of tunneling systems
with almost identical energy splitting and exhibiting a correlated
tunneling motion with an effective tunneling loop. It is unclear at this
point how large the fraction of tunneling systems in glasses is, having
the required properties.

As stated above, the pair model aims to describe the dielectric
susceptibility data and no specific predictions were made for polarization
echo experiments.\cite{Wue02} Nevertheless, one would certainly expect a
pronounced frequency dependence since the parameter~${\it\Gamma}$ is
inversely proportional to the measuring frequency.
\bigskip

\centerline{\it 3. Nuclear Quadrupole Model}
\bigskip

Very recently, an alternative explanation for the magnetic field effects
of the dielectric properties of insulating glasses has been
suggested.\cite{Wue02b} The central assumption of this model is, that the
nuclear properties of the tunneling systems are responsible for the
observed phenomena. Nuclei with a spin~$I\ge 1$ also carry a nuclear
quadrupole moment~$Q$. Because of electric field gradients present in
glasses, both, the ground state and the excited state of tunneling systems
containing nuclei with a quadrupole moment, will exhibit a quadrupole
splitting~$\hbar\omega_{\rm Q}$, that is generally different for the two
levels. Magnetic fields couple to the nuclear magnetic dipole moment and
lead to a nuclear Zeeman splitting $\hbar\omega_{\rm L}=g\mu_{\rm N}B$.
Here $g$ represents the $g$-factor of the nuclei and $\mu_{\rm N}$ the
nuclear magneton. The behavior of such systems in polarization echo
experiments depends on the relative magnitude of the two energies. At low
magnetic fields ($\hbar\omega_{\rm L}\ll \hbar\omega_{\rm Q}$) the
amplitude of a two-pulse polarization echo for nuclei with half-integer
spin and $I\geq 3/2$ should approximately be given by \cite{Wue02b}
\begin{eqnarray}
&&A(t_{12},\omega_{\rm L})= A(t_{12},\infty )\bigg\{ 1-a\nonumber\\
&&\;+\;b_{1/2}\!\int\limits_0^1\! p(x)\cos\left[\sqrt{x^2 +
(I+1/2)^2(1-x^2)}\,\,\omega_{\rm L}t_{12}\right ]{\rm d}x
\nonumber\\
&&\;+\!\sum_{\beta >1/2} b_ \beta \!\int\limits_0^1\!\!
p(x)\cos\,(2\beta\omega_{\rm L}t_{12}x)\,{\rm d}x \bigg\}
\label{weak_fields}
\end{eqnarray}
\noindent where $a$ and $b_\beta$ are coefficients depending on the
nuclear spin, and $A(t_{12},\infty )$ denotes the echo amplitude in the
limit of high magnetic fields. The parameter~$x$ represents the
projection of the quantization axis of the magnetic moments onto the
quadrupole axis and $p(x)$ is its normalized distribution. The sum runs
over $\beta=1/2, \dots, I$ for half-integer spins.\cite{footnote1} For
\hbox{$B=0$} the amplitude of the echo is given by
\hbox{$A(t_{12})=A(t_{12},\infty )[1-a + \sum_\beta b_\beta]$}. Compared
to the amplitude caused by an ensemble of simple two-level systems it is
reduced by the fraction $(a - \sum_\beta b_\beta)$, which arises from
the interference of different quadrupole levels.\cite{Ebe80} At small
magnetic fields the degeneracy of the nuclear magnetic levels is lifted
and causes an additional line broadening due to additional transition
energies. The non-monotonic variation of the echo amplitude with the
magnetic field results from the magnetic field dependence of the
distribution of nuclear spin levels and the interference of their
contributions to the signal. The first minimum of (\ref{weak_fields}) is
expected to occur roughly at $\omega_{\rm L}t_{12}\approx
0.5\pi$.\cite{footnote2} Inserting the expression for the nuclear Zeeman
energy this condition leads to
\begin{equation}
B_{\rm min}t_{12}\approx{0.5\,\pi\hbar\over g \mu_{\rm N}}\;.
\label{minimum}
\end{equation}

\noindent Thus the nuclear quadrupole model predicts that the position
of the minimum only depends on the nuclear magnetic dipole moment of the
tunneling particle. As a consequence echo experiments are expected to
reveal interesting information on the material dependence of the
magnetic field phenomena in glasses. It is noteworthy, however, that the
tunneling systems in glasses are probably not caused by the motion of
single atoms, but are likely to involve the correlated motion of small
atomic clusters. In this case, the interpretation of the non-monotonic
variation of the echo amplitude with magnetic field is obviously more
complex.

At large magnetic fields ($\hbar\omega_{\rm L}\gg \hbar\omega_{\rm Q}$)
the situation simplifies significantly and reduces to the case of a simple
two-level system, because the nuclear Zeeman splitting leads to equally
spaced spin levels for both, the ground states and the excited states of
the tunneling systems. According to the nuclear quadrupole
model\cite{Wue02b} the variation of the two-pulse echo amplitude at large
magnetic fields should approximately be given by
\begin{equation}
{A(B)\over A(B\to \infty)}\approx 1-{B_0^2\over B^{2}}\;,
\label{strong_fields}
\end{equation}

\noindent where the constant~$B_0$ marks the magnetic field at which
$\hbar\omega_{\rm L}\approx \hbar\omega_{\rm Q}$. Note that at high
fields the normalized amplitude does not depend on the delay time
$t_{12}$ whereas at low fields a delay time dependence is expected. From
the magnetic field strength of the transition from the low-field to the
high-field regime it is possible to estimate the magnitude of the
nuclear quadrupole splitting with the help of this model.

\section{Experimental Technique}

The experiments were performed in a $^3$He-$^4$He dilution refrigerator in
order to reach temperatures down to about 10\,mK, which are necessary to
study coherent polarization echoes in glasses. The cooling power of the
apparatus was sufficient to overcome the heat input by the two coaxial
cables carrying the r.f.~signals and by the r.f.~power dissipated in the
experiment itself. All r.f.~lines inside the refrigerator were semi-rigid
coaxial cables with low insertion loss up to the GHz range. Inside the
vacuum enclosure of the refrigerator the coaxial cables are made from
superconducting niobium in order to keep the heat input by thermal
conduction as low as possible. Thermalization of the niobium cables is
performed at the still with home made strip-line devices on a
single-crystal sapphire substrate. The temperature was measured by an
Au:Er susceptibility thermometer and a carbon resistor, both attached to
the mixing chamber of the apparatus.
\bigskip

\centerline{\it 1. Echo Generation and Detection}
\bigskip

The echo experiments were performed in a re-entrant microwave resonator
attached to the mixing chamber of the dilution refrigerator. The resonator
was made from gold plated oxygen free copper. The sample discs with a
thickness of about 0.5$\,$mm and a diameter of 8$\,$mm were mounted in the
region between the center post and the bottom of the resonator. In this
area a large portion of the electric field component of the resonator is
concentrated and, because of the diameter (8$\,$mm) of the center post,
the field was rather homogeneous. The r.f.~signal was inductively coupled
into the resonator by a loop which could be rotated in order to adjust the
degree of coupling. Two modes, the ground mode at 1\,GHz and the first
overtone at 4.6\,GHz were used in our experiments.

The echo signal was picked up by a second loop in the resonator and was
pre-amplified by approximately 43\,dB using a helium-cooled microwave
amplifier located in the helium bath outside the vacuum chamber. The
signal was mixed outside the cryostat with a reference signal with the
same frequency. The demodulated signal was further amplified by a video
amplifier and digitized by a fast digital oscilloscope. Depending on the
experiment 1000 to 10000 subsequent echo cycles were digitally averaged in
order to reach an acceptable signal to noise ratio.

The r.f.~signal was generated by a microwave synthesizer, split into two
parts and fed into the receiver and transmitter branch. The latter
consisted of a variable attenuator, a phase shifter, and a {\it p-i-n}
diode switch used to shape the microwave pulses with an on/off transit
time of about 15$\,$ns. The TTL control signals of the diode switch were
provided by a digital pulse and delay generator.

Phase sensitive detection has two advantages for our experiment: Firstly,
sensitivity is increased by 3$\,$dB because all noise in the quadrature
component of the signal is suppressed. Secondly, the output of a phase
sensitive detector is bipolar, meaning that the mean value of noise of the
detector signal vanishes. In a conventional rectifying (\lq square law')
detector the noise output is unipolar and will therefore average to a
finite value. Furthermore, mixing of the bipolar random noise signal with
the fixed-polarity echo signal in a rectifying detector would lead to a
nonlinear amplitude dependence of the averaged echo at low signal levels.
\bigskip

\centerline{\it 2. Magnetic Field Generation}
\bigskip

In most experiments discussed, here the magnetic field was
generated by a home-made superconducting magnet consisting of
about 1000\,m NbTi wire wound around a stainless steel cylinder
with a 22\,mm drill-hole and 64\,mm in length. The wire was fixed
with a thick layer of epoxy (Stycast 2850 FT). The coil was
attached to the mixing chamber of the dilution refrigerator and
the sample was placed in the middle of the cylinder. The
homogeneity of the magnetic field was estimated to be better than
$5\times10^{-3}$. The maximum current through the coil was about
2\,A corresponding to a magnetic field strength of 0.46\,T. The
operation of the magnet at constant field had no noticeable
influence on the performance of the cryostat. However, whenever
the magnetic field was altered, the temperature of the sample
holder increased significantly due to eddy currents, and, after
reaching a constant magnetic field, it was slowly relaxing back to
the mixing chamber temperature. For experiments at higher fields
we used a standard superconducting solenoid with a field range up
to 5\,T, which was installed at the mixing chamber of our
cryostat.\bigskip

\centerline{\it 3. Samples}
\bigskip

Four different insulating glasses were investigated, namely the two
commercially available multi-component glasses BK7 and Duran, the
multi-component glass \hbox{a-BaO-Al$_2$O$_3$-SiO$_2$}~(BAS), and vitreous
silica containing 1200\,ppm OH$^-$ (Suprasil~I). The main components of
these glasses are listed in Table~\ref{table1} as determined by atomic
emission spectroscopy.\cite{Str00f} In addition, in Table~\ref{table2} we
have listed for the relevant elements, the nuclear spins, the quadrupole
moments of the isotopes with non-zero quadruple moment, and the natural
abundance of these isotopes.
\begin{table}[h]
\caption[]{Mole fraction of the constitutes of the four glass samples
BAS, BK7, Duran and Suprasil\,I as determined by atomic emission
spectroscopy.~\cite{Str00f}}
\renewcommand{\arraystretch}{1.2}
\setlength\tabcolsep{7.2pt}
\begin{tabular}{@{}llccccc}
\hline\noalign{\smallskip}
&&& BAS & BK7 & Duran & Suprasil\,I \\
\hline \noalign{\smallskip}
SiO$_2$ &$\!\!$(\%)    &&  72.7   &  74.8      & 83.4     & 100  \\
B$_2$O$_3$ &$\!\!$(\%) &&  0.72   &   9.6      & 11.6     &   0  \\
Al$_2$O$_3$ &$\!\!$(\%)&&  8.8    &   0.03     & 1.14     &   0  \\
Na$_2$O &$\!\!$(\%)    &&  0.28   &   10.1     & 3.4      &   0  \\
K$_2$O &$\!\!$(\%)     &&  0.064  &   4.7      & 0.41     &   0  \\
BaO     &$\!\!$(\%)    &&  17.0   &   0.76     & 0.005    &   0  \\
\hline
\end{tabular}
\label{table1}
\end{table}

\begin{table}[h] \caption[]{Natural abundance, nuclear spin $I$ and
nuclear quadrupole moment $Q$ of isotopes with $I>1/2$ present in the
glass samples.}
\renewcommand{\arraystretch}{1.2}
\setlength\tabcolsep{9pt}
\begin{tabular}{@{}lcccc}
\hline\noalign{\smallskip}
& abundance (\%) & $I$ &  $g$ & $Q$ (barn) \\
\hline \noalign{\smallskip}
$^{10}$B   &  19.9   &  3   & 0.60    & 0.08       \\
$^{11}$B   &  80.1   &  3/2 & 1.79    & 0.04       \\
$^{17}$O   &  0.05   &  5/2 & $-0.76$ & $-0.03$    \\
$^{27}$Al  &  100    &  5/2 & 1.46    & 0.14       \\
$^{23}$Na  &  100    &  3/2 & 1.48    & 0.11       \\
$^{39}$K   &  93.3   &  3/2 & 0.26    & $0.06$     \\
$^{41}$K   &  6.7    &  3/2 & 0.14    & $0.07$     \\
$^{135}$Ba & 6.59    &  3/2 & 0.56    & $0.16$     \\
$^{137}$Ba & 11.23   &  3/2 & 0.62    & $0.25$     \\
\hline
\end{tabular}
\label{table2}
\end{table}

\begin{figure}[b]
\includegraphics[width=0.78\linewidth]{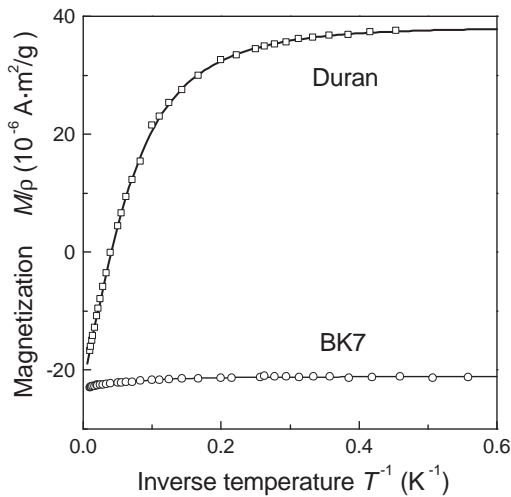}
\vskip -2 mm \caption{Magnetization of BK7 and Duran as a function
of the inverse temperature. The measurements were carried out in a
field of 5.5\,T.}\label{Fig1}
\end{figure}

Since we were concerned about effects caused by magnetic
impurities present in the samples we performed magnetization
measurements in a SQUID magnetometer to determine the amount of
magnetic impurities. Fig.~\ref{Fig1} shows the result of such
measurements on BK7 and Duran. Although these two glasses have a
very similar composition, the change in magnetization and thus the
amount of magnetic impurities differ by roughly a factor of
twenty. We conclude from this measurement that the BK7 sample
contained about 5\,ppm of magnetic impurities with a g-factor of
$g\approx 2$ and a total spin of $J=5/2$, while in Duran about
120\,ppm magnetic impurities were found with $g\approx 2$ and
$J\approx 5/2$.

\section{Experimental Results and Discussion}

We have studied the influence of magnetic fields on the amplitude of
coherent polarization echoes depending on various parameters such as
frequency, delay time, electric field strength and temperature. As we will
see, the experimental results allow to distinguish between the different
models discussed above.
\bigskip\bigskip

\centerline{\bf A. Spontaneous Echoes in Magnetic Fields}
\bigskip

\centerline{\it 1. Electric Field Dependence}
\bigskip

Let us first discuss the dependence of the echo amplitude on the electric
field strength of the exciting microwave pulses. In all experiments the
widths of the pulses were kept constant, being $t_{\rm p1}\approx 100\,$ns
and \hbox{$t_{\rm p2}\approx 200\,$ns}, the electric field strength was
the same for both pulses. At small electric fields the amplitude of the
echo is expected to rise proportional to $F_0^3$. It reaches a maximum at
a field fulfilling the condition ${\it\Omega}t_{\rm p2}\approx\pi$ for the
second pulse. At higher electric fields the shape of the echo starts to
change, splitting eventually into different negative and positive
parts.\cite{Gur90} In this regime it is not obvious which maximum one
should take as the echo amplitude. Therefore, we have decided to plot
always the integrated echo amplitude, because this is a well-defined
quantity even at large fields. After passing through the maximum the
integrated echo amplitude is expected to decrease towards higher fields,
since in this range the signal consists of negative and positive
components.

\begin{figure}[t]
\vskip - 8mm
\includegraphics[width=1.00\linewidth]{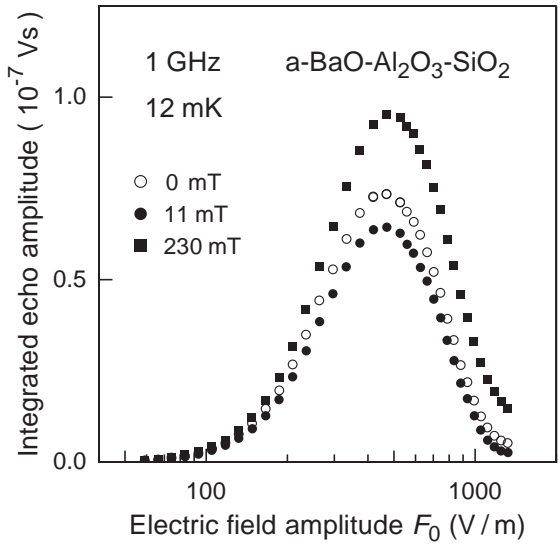}
\vskip - 7mm \caption{Electric field dependence of the amplitude
of the spontaneous polarization echo in BAS at zero field, 11\,mT
and 230\,mT (from reference~8).}\label{Fig2}
\end{figure}

In Fig.~\ref{Fig2} the integrated amplitude of two-pulse polarization
echoes generated in BAS at different magnetic fields is plotted as a
function of the applied electric field strength. All curves show the
expected variation with the electric field strength, but the absolute
value of the echo amplitude is different at different magnetic fields.
It is noteworthy that the echo amplitude is a non-monotonic function of
the applied magnetic field. The fact, that independent of the applied
magnetic field the maximum of the integrated echo amplitude is always
found at the same electric field strength of about 450\,V/m indicates
that the mean effective Rabi frequency remains virtually unaltered in
magnetic fields. From this we conclude that neither the dipole moment
$p$ nor the tunneling splitting~${\it\Delta}_0$ depend on the applied
magnetic field. This observation disfavors the model proposed by
Kettemann et al.~\cite{Ket99} as an explanation for the magnetic field
dependence of the echo amplitude.

\bigskip

\centerline{\it 2. Material Dependence}
\bigskip

\begin{figure}[t]
\includegraphics[width=0.84\linewidth]{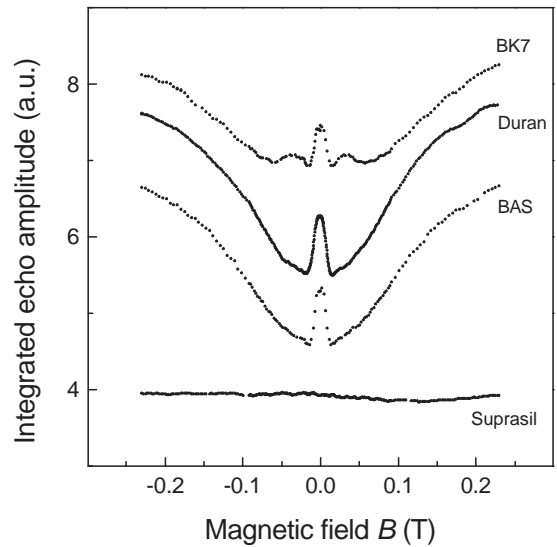}
\vskip-2mm \caption{Magnetic field dependence of the amplitude of
spontaneous polarization echoes in BK7, BAS, Duran, and
Suprasil~I. All data were taken at 12\,mK, \hbox{$t_{12}=
2\,\smu$s} and roughly 1\,GHz, except that for Duran, where the
delay time was $t_{12}= 1.7\,\smu$s.} \label{Fig3}\end{figure}

We have studied the coherent properties of four different glasses in
magnetic fields in order to investigate the material dependence of the
magnetic field effects. The amplitude of two-pulse polarization echoes
of these four glasses is shown in Fig.~\ref{Fig3} as a function of
magnetic field.\cite{footnote3} In contrast to many other
low-temperature properties of glasses the influence of magnetic fields
on the amplitude of spontaneous echoes is obviously not universal.
Several conclusions can be drawn from the results shown in
Fig.~\ref{Fig3}. The occurrence and the magnitude of the magnetic field
dependence is independent of the concentration of magnetic impurities
present in the glass samples. In particular, BK7 and Duran show similar
effects although the concentration of magnetic impurities differs by at
least a factor of 20 as demonstrated by the magnetization measurements
shown in Fig.~\ref{Fig1}.

Perhaps the most remarkable result of the measurements plotted in
Fig.~\ref{Fig3} is the fact that Suprasil~I shows no magnetic field effect
within the accuracy of our experiment. The slight difference between
positive and negative fields is caused by drift problems in the
experiment. Whereas Duran, BAS and BK7 contain isotopes with non-zero
nuclear quadrupole moment, Suprasil is virtually free of such
isotopes.\cite{footnote4} This strongly indicates that nuclear quadrupoles
play a crucial role in the magnetic field effects, and therefore, our
result clearly favors the nuclear quadrupole model.\cite{Wue02b}

\begin{figure}[b]
\includegraphics[width=0.83\linewidth]{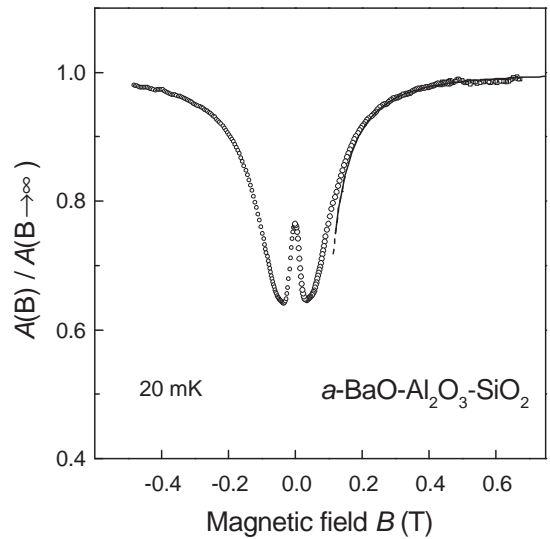}
\vskip -3 mm\caption{Amplitude of spontaneous polarization echoes
in BAS in magnetic fields up to 0.7\,T. The data were taken in at
1\,GHz, 20\,mK and $t_{12}= 1\,\smu$s. The solid line represents
the expected variation at high fields according to the quadrupole
model.}\label{Fig4}\end{figure}

The variation of the echo amplitude with applied magnetic field is
similar for Duran, BK7 and BAS, but not identical. All three
samples exhibit a central maximum at $B=0$. At high fields the
amplitude of the echo rises above its value at zero magnetic field
and seems to reach saturation at fields above 200\,mT.
Fig.~\ref{Fig4} shows the variation of the echo amplitude for BAS
at $t_{12}=1\umu$s in an extended field range. In this plot the
saturation effect becomes evident. According to the nuclear
quadrupole model this saturation occurs at magnetic fields at
which the nuclear Zeeman energy exceeds the mean quadrupole
splitting. This condition seems to be met roughly at the same
magnetic field in all three glasses (see Fig.~\ref{Fig3}),
although the relevant nuclei are probably either $^{11}$B or
$^{27}$Al with different nuclear quadrupole moments.

The solid line in Fig.~\ref{Fig4} represents the
prediction~(\ref{strong_fields}) of the quadrupole model for large
fields with $B_0 = 60$\,mT as fitting parameter. It describes the high
field limit almost perfectly. From the value of~$B_0$ we deduce the
Lamor frequency~$\omega_{\rm L} = 4.2$\,MHz for the aluminium nuclei.
Assuming further that $\hbar\omega_{\rm L}\approx \hbar\omega_{\rm Q}$
at~$B_0$, we obtain for the electric field gradient the value~$8\times
10^{20}$\,Vm$^{-2}$ which is of the right order of magnitude.

\bigskip
\centerline{\it 3. Delay Time Dependence}
\bigskip

\begin{figure}[b]
\includegraphics[width=0.83\linewidth]{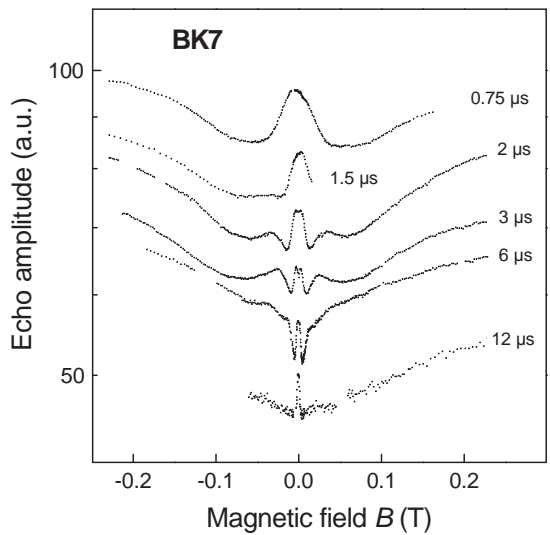}
\vskip -3 mm
\caption{Amplitude of two-pulse echoes generated in
BK7 as a function of magnetic field at different delay times. All
data sets were taken at 4.6\,GHz and 12\,mK except that for
$t_{12}=2\,\smu$s which was taken at
0.9\,GHz.}\label{Fig5}\end{figure}

Further remarkable observations were made in measurements of the magnetic
field dependence of the amplitude of two-pulse echoes with different delay
times~$t_{12}$ between the exciting pulses. As an example, we show in
Fig.~\ref{Fig5} the amplitude of spontaneous echoes excited in BK7 as a
function of the applied magnetic field for different delay times. The
pattern of the magnetic field dependence is clearly different for
different values of~$t_{12}$. This observation indicates that the applied
magnetic field influences the free phase development of the tunneling
systems between and after the two excitation pulses.

\begin{figure}[t]
\includegraphics[width=0.82\linewidth]{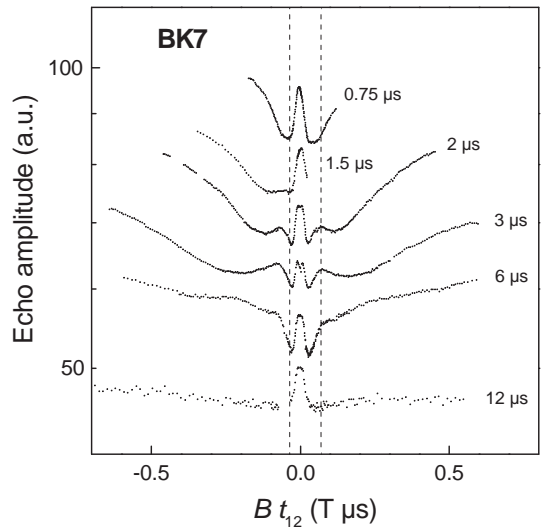}
\vskip -3 mm\caption{Amplitude of two-pulse echoes generated in
BK7 as a function of the product of the magnetic field~$B$ times
the delay time $t_{12}$. The dashed lines mark the positions of
typical features of the echo amplitude.} \label{Fig6}\end{figure}

This can be made even more obvious by plotting the amplitude as a
function of the product of the magnetic field and the delay time~$t_{\rm
12}$. Fig.~\ref{Fig6} shows such a plot for the data displayed in
Fig.~\ref{Fig5}. It is remarkable that several features of the curves at
different delay times coincide in this representation. In particular,
the width of the central maximum in this plot is nearly identical in all
cases. In addition, we have marked two features by dashed lines, the
minimum at negative fields and the second maximum at positive field
strength. Such an invariance of the pattern with respect to the product
$Bt_{\rm 12}$ is expected in the quadrupole coupling model at small
magnetic fields. According to this approach, mixing between nuclear
levels is the cause of these features. From the data shown in
Fig.~\ref{Fig3}, the first minimum is observed at about 14\,mT in all
three glasses. In BK7 the two boron isotopes~$^{10}$B and $^{11}$B as
well as sodium are likely to cause the magnetic field effect
(see~Table~\ref{table2}). In the case of BAS the main contribution is
probably due to $^{27}$Al~nuclei. Using the rough
approximation~(\ref{minimum}) and the appropriate $g$-factors we deduce
values of about 11\,mT for the minima in fair agreement with our
observations. Applying~(\ref{minimum}) to the lighter boron
isotope~$^{10}$B leads to a greater value of the magnetic field at the
minimum. However, it is not clear whether~(\ref{minimum}) is applicable
since a theory for the contribution of integer spins is still lacking.

The absence of a second maximum in case of Duran and BAS
(Fig.~\ref{Fig3}) is another interesting observation made in our
experiments. In the light of the quadrupole coupling model this
might be caused by the larger amount of magnetic impurities in
these materials compared to BK7, because it is conceivable that
electronic spins lead to a distribution of local magnetic fields
and therefore might wash out certain interference patterns.
\goodbreak

Despite the good qualitative and partially even quantitative agreement
of the quadrupole coupling model we want to point out that not all
details of the curves shown in Fig.~\ref{Fig5} can be described by
(\ref{weak_fields}).
\bigskip

\centerline{\it 4. Frequency Dependence}
\bigskip

\begin{figure}[t]
\includegraphics[width=0.82\linewidth]{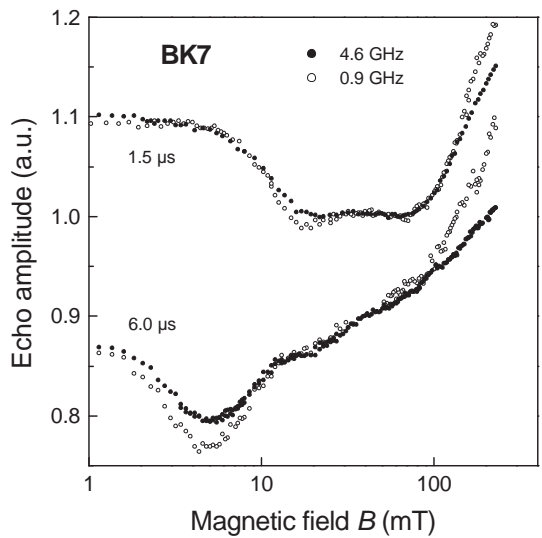}
\caption{Amplitude of two-pulse echoes generated in BK7 as a
function of magnetic field at two different frequencies and at two
different delay times.} \label{Fig7}\end{figure}

A pronounced frequency dependence of the magnetic field effects in
polarization echo experiments is expected from both models assuming a
direct coupling of magnetic fields to tunneling
systems.\cite{Ket99,Wue02} Fig.~\ref{Fig7} shows the magnetic field
dependence of the echo amplitude of BK7 measured at 0.9\,GHz and
4.6\,GHz, and at two different delay times. Note that the magnetic field
is plotted on a logarithmic scale. Again, the curves at different delay
times exhibit a different pattern as a function of the applied magnetic
field, but hardly any frequency dependence is seen within the accuracy
of the experiment, although frequencies differ by a factor of five. The
remaining differences between the curves in particular at the long delay
time of $6\mu$s and at higher magnetic fields could be caused by
experimental problems.

The apparent absence of a frequency dependence in this experiment
provides further evidence that the models favoring a direct coupling of
magnetic fields to tunneling systems are not applicable in this case. In
contrast, no (strong) variation with measuring frequency is expected
within the nuclear quadrupole coupling model, because the relevant
energy scales are given by the nuclear properties of the tunneling
systems.
\bigskip

\centerline{\it 5. Temperature Dependence}
\bigskip

\begin{figure}[h]
\includegraphics[width=0.82\linewidth]{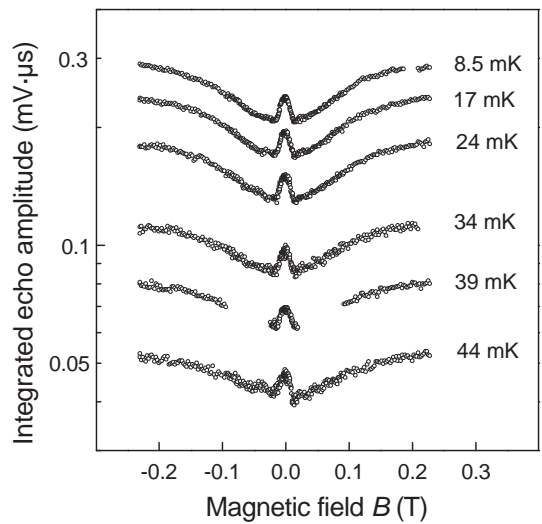}
\caption{Amplitude of two-pulse echoes generated in Duran as a function
of magnetic field at different temperatures. In this experiment the
microwave frequency was roughly 1~GHz and the delay time 1.7\,$\smu$s.}
\label{Fig8}\end{figure}

If the nuclear quadrupole model is the correct description, the magnetic
field effects observed in the two-pulse echo experiments are expected to
be temperature independent because of their pure quantum nature. We have
studied the temperature dependence of the magnetic field effects of two
glasses: Duran and BAS. The result for Duran is shown in Fig.~\ref{Fig8}
where the integrated echo amplitude is plotted on a logarithmic scale.
With rising temperature the amplitude of the echo decreases strongly.
However, the pattern of the magnetic field variation remains nearly
unaltered. There is a small decrease of the magnetic field variation
relative to the absolute echo amplitude, which is hardly visible in this
representation.

\begin{figure}[b]
\includegraphics[width=0.82\linewidth]{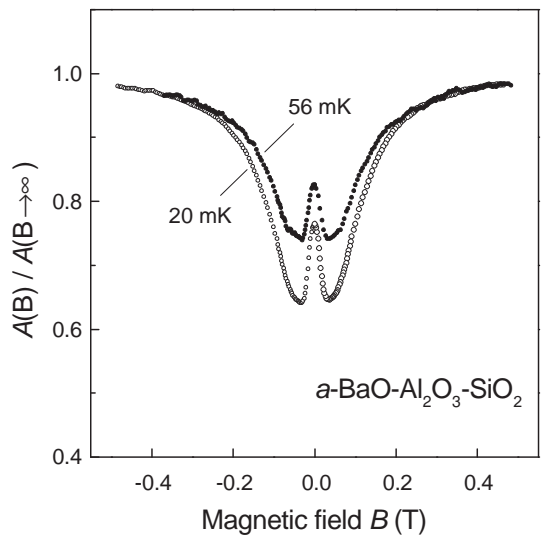}
\caption{Amplitude of two-pulse echoes generated in BAS as a function of
magnetic field at two different temperatures. The measurement was
performed about 1~GHz and a delay time of 1\,$\smu$s. In both cases the
echo amplitude is normalized to its value at high fields.}
\label{Fig9}\end{figure}

The magnetic field dependence of the echo amplitude in BAS at 20\,mK and
56\,mK is shown in Fig.~\ref{Fig9} where the normalized
amplitude~$A(B)/A(B\to\infty)$ is plotted on a linear scale. As in the
case of Duran, the overall pattern remains unchanged as the temperature
increases although the relative variation is larger at the lower
temperature. With rising temperature the magnetic field effect vanishes
obviously more rapidly than the amplitude of the echo itself. At a first
glance this is unexpected from the point of view of the quadrupole
model. In an attempt to explain this finding we may distinguish between
tunneling particles with and without nuclear quadrupole moment. In both
cases, the temperature dependence of the echo amplitude is assumed to be
independent of magnetic fields. However, it is conceivable that the
temperature dependence of the dephasing time of systems carrying a
nuclear quadrupole moment is different from that of systems without a
nuclear quadrupole moment. In this case the magnitude of the magnetic
field effect would also depend on temperature.

Finally we want to add that the decay of spontaneous echoes in glasses is
much faster than expected.\cite{Ens02b} According to theory\cite{Bla77}
the decay is due to spectral diffusion and is consequently determined by
the thermal relaxation of tunneling systems with an energy
splitting~$E\approx 2 k_{\rm B}T$. In this context one might speculate
whether a novel, hitherto unconsidered mechanism gives rise to fast
relaxation because of the more complex level scheme of tunneling systems
with a nuclear quadrupole moment.

\section{Summary and Outlook}

We have measured the amplitude of the two-pulse echoes generated in four
different glasses in magnetic fields. Three glasses exhibit a striking
non-monotonic magnetic field variation, but vitreous silica Suprasil~I
shows no magnetic field effect. From the absence of this phenomenon in
Suprasil~I, and from the influence of other parameters, such as delay
time, electric field, temperature, and frequency on the magnetic field
effect we conclude that tunneling particles carrying a nuclear quadrupole
moment are responsible for the surprising phenomena. The nuclear
quadrupole of these systems leads to a splitting of the nuclear spin
levels even at zero magnetic field, which is different for the ground
state and the excited state. The presence of this multi-level structure
reduces the echo amplitude compared to that of simple two-level systems.
As long as the Zeeman energy is smaller than the mean quadrupole
splitting, mixing between the nuclear spin levels leads to a non-monotonic
variation of the echo amplitude with the magnetic field. At a large
magnetic field the Zeeman effect is dominant and the echo amplitude
saturates at the value expected for simple two-level systems.

We believe that nuclear quadrupoles also play an important role in other
experiments on glasses at low temperatures. For example the magnetic field
dependence of the dielectric susceptibility at low frequencies is likely
to be caused by relaxation effects involving nuclear spin levels. In
addition, we like to point out that the fact that the nuclear properties
of the tunneling particles matter, might allow to study the microscopic
nature of the tunneling systems in glasses by investigating glasses of
selected composition.

\section*{Acknowledgment}

We thank R. Weis, A. W\"urger, P. Strehlow  for many stimulating
discussions. In particular, we acknowledge the help of B. Renker in the
measurement of the impurity concentration of BK7 and Duran. This work was
supported by the Deutsche Forschungsgemeinschaft.



\begin{references}
\bibitem{footnote1}It should be noted that the best fit to the experimental
data was obtained in reference~\cite{Wue02b} with $p(x)=3x^2$, and that
until now no theory has been worked out for nuclei with integer spin.
\bibitem{footnote2}In the nuclear quadrupole model, this problem has been
worked out in more detail. In fact, the structure of the minimum is more
complex because of the contribution of the different Zeeman levels. In
addition, the exact position of the minimum depends on the distribution
function~$p(x)$.
\bibitem{footnote3}If not stated differently, the echo amplitudes were multiplied
by different arbitrary factors in order to obtain a unified
representation of the data.
\bibitem{footnote4}The only isotope with non-zero nuclear quadrupole in pure Suprasil~I is
               $^{17}$O ($I= 5/2$) which has an natural abundance of only 0.05\%.
\end{references}
\end{document}